\documentclass[12pt]{iopart}

\usepackage{graphicx,caption,hyperref,cite,tabularray,indentfirst}
\usepackage{xcolor}
\usepackage{soul}

\begin{document}

\title
{Diagnosing electric and magnetic fields in laser-driven coil targets}

\author{Yang Zhang\textsuperscript{1,2,*}, Lan Gao\textsuperscript{3, *}, Hantao Ji\textsuperscript{1,3}, Brandon K. Russell\textsuperscript{1}, Geoffrey Pomraning\textsuperscript{1}, Jesse~Griff-McMahon\textsuperscript{1}, Sallee~Klein\textsuperscript{4}, Carolyn~Kuranz\textsuperscript{4}, Mingsheng~Wei\textsuperscript{5}}

\address{1. Department of Astrophysical Sciences, Princeton University, Princeton, NJ, 08544, USA} 
\address{2. University Corporation for Atmospheric Research, Boulder, CO, 80301, USA }
\address{3. Princeton Plasma Physics Laboratory, Princeton University, Princeton, NJ, 08543, USA}
\address{4. University of Michigan, Ann Arbor, Michigan, 48109, USA}
\address{5. Laboratory for Laser Energetics, University of Rochester, Rochester, New York, 14623, USA}

\ead{yz0172@princeton.edu}
\ead{lgao@pppl.gov}
\vspace{10pt}

\begin{abstract}

Laser-driven capacitor coils are widely used to generate intense magnetic fields for various applications in high-energy-density physics research. Accurate measurement of the magnetic fields is essential but challenging, due to the overlapping contributions from magnetic and electric fields in proton radiography, which is the primary tool diagnosing the field generation around the coils. In this study, we systematically analyze proton radiographs obtained from laser-driven capacitor-coil targets along two orthogonal axes under various electromagnetic field conditions, including magnetic field only, electric field only, and combined electromagnetic fields. By analyzing key features in the radiographs, we distinguish and characterize the respective contributions from magnetic and electric fields. Using detailed simulations validated by experimental benchmarks, methods to isolate and quantify the magnetic field and electric field are given. The methods are successfully applied to determine the electric current and charge distribution in a double coil configuration. 
Our findings provide insights into improving the diagnostic capability of proton radiography, potentially leading to more accurate measurements of electromagnetic fields and enhancing the utility of laser-driven capacitor coils in high-energy-density experiments.

\end{abstract}

%
%
%
%
%

\section{Introduction}
A laser-driven capacitor coil is a highly effective device for generating intense magnetic fields\cite{daido1986generation,courtois2005creation,santos2015laser,law2016direct,gao2016ultrafast,zhang2018generation,morita2023generation}. It consists of two parallel plates connected by a conductive wire or coil \cite{gao2016ultrafast,morita2023generation}. When a high-intensity laser irradiates one of the plates, it produces superthermal electrons during the intense laser-solid interaction that are subsequently collected by the opposing plate, resulting in a voltage difference between two plates.  This voltage drives an electric current through the coil that can be as large as hundreds of kiloamperes, resulting in the generation of strong magnetic fields.

The ability to generate such strong, transient magnetic fields has positioned laser-driven capacitor coils as an essential tool in high-energy-density (HED) physics research, enabling detailed studies of fundamental plasma processes in magnetized HED experiments, and associated particle dynamics under extreme physical conditions. Their applications span a wide range of areas, including laboratory astrophysics \cite{chien2019study,chien2023non,zhang2023ion,yuan2023push,ji2024study}, inertial confinement fusion\cite{sakata2018magnetized,matsuo2021enhancement,pisarczyk2022influence}, and charged particle acceleration \cite{morita2020application,Bailly_NC_2018}.

In addition to the magnetic fields generated by the large current flowing through the coil, laser-driven systems often produce significant electric fields due to charge accumulation along the conductive path of the coil. The interplay between these electric and magnetic fields creates highly complex electromagnetic field distributions. Therefore, understanding the spatiotemporal evolution of these fields is critical for optimizing the design of capacitor coils and enhancing their utility in experimental applications. 


Proton radiography\cite{schaeffer2023proton}, also known as proton deflectometry, has emerged as a primary diagnostic technique for measuring the electromagnetic field distributions in such systems\cite{gao2016ultrafast,fiksel2016simple,peebles2020axial,bradford2020proton,bradford2021measuring,yuan2021full,vlachos2024laser}. This technique utilizes high-energy protons (in the MeV range) that are directed through the target region. These protons are deflected by electromagnetic fields via the Lorentz force, creating deflection patterns that are captured on detectors, e.g., CR-39 or radiochromic film (RCF). 
These patterns provide path-integrated information about the electromagnetic fields within the target area. 
Since deflections can be caused by both magnetic and electric fields\cite{peebles2022assessment,sutcliffe2021new}, determining their individual contributions from a proton radiograph requires a clear understanding of the distinct patterns produced by each field.

Previous studies have utilized face-on void (or bubble) size in proton radiographs to infer coil current and the associated magnetic field strength \cite{gao2016ultrafast,peebles2022assessment,bradford2020proton}. Since such voids can be produced either by a positively charged wire (via electric fields) or by current flowing through the wire (via magnetic fields), a single void size from one radiograph alone cannot unambiguously distinguish between the two. 

Because electric and magnetic deflections exhibit different scaling with proton energy, for magnetic fields, the void size scales as ${E}_p^{-1/4}$\cite{gao2016ultrafast}, whereas for electric fields it scales as ${E}_p^{-1/2}$ \cite{bradford2021measuring}. A dual-energy proton deflection method has been proposed to differentiate the field types \cite{sutcliffe2021new}. While promising, this approach relies on the availability of a broad proton energy spectrum. The combination of high energy and wide energy spread—may not be readily achievable at all experimental facilities\cite{peebles2022assessment}. 

An alternative is to reverse the proton beam direction in a second shot, flipping the magnetic deflection while keeping the electric deflection unchanged. Vlachos \textit{et al.} \cite{vlachos2024laser} demonstrated this for a magnetic-dominated case. This method requires two separate shots for a single measurement, and its reliability depends on maintaining low shot-to-shot variability.

Side-on grid rotation analysis offers another approach to infer coil current and magnetic field strength in capacitor-coil targets  \cite{peebles2020axial,bradford2020proton,bradford2021measuring,peebles2022assessment}. Since the induced rotation angle depends solely on the magnetic field, this technique provides a direct and unambiguous measurement of magnetic field strength. As the proton beam is aligned with the main magnetic field direction in this configuration, the resulting deflections are smaller, leading to reduced field sensitivity.


In this study, we conduct a systematic investigation of proton radiograph patterns produced by a laser-driven coil target, focusing on a case where the magnetic field originates from a coil current and the electric field arises from a uniform charge distribution along the coil. Using ray tracing simulations, we analyze proton deflection patterns along two orthogonal axes—face-on and side-on—and examine the respective contributions of magnetic fields, electric fields, and their combined effects. By identifying and characterizing key features in the deflection patterns for each field configuration, we develop an approach to estimate the magnetic field and electric field contributions from proton radiographs. We then apply this methodology to our experimental data and find strong consistency between our analysis and measured results\cite{gao2025record}. Our study provides insights into how proton radiographs can be interpreted in laser-driven coil targets, contributing to a more refined understanding of field characterization in high-energy-density physics experiments.

\section{Simulation Setup}
\begin{figure*}[h]
\centering
\includegraphics[width=1\textwidth]{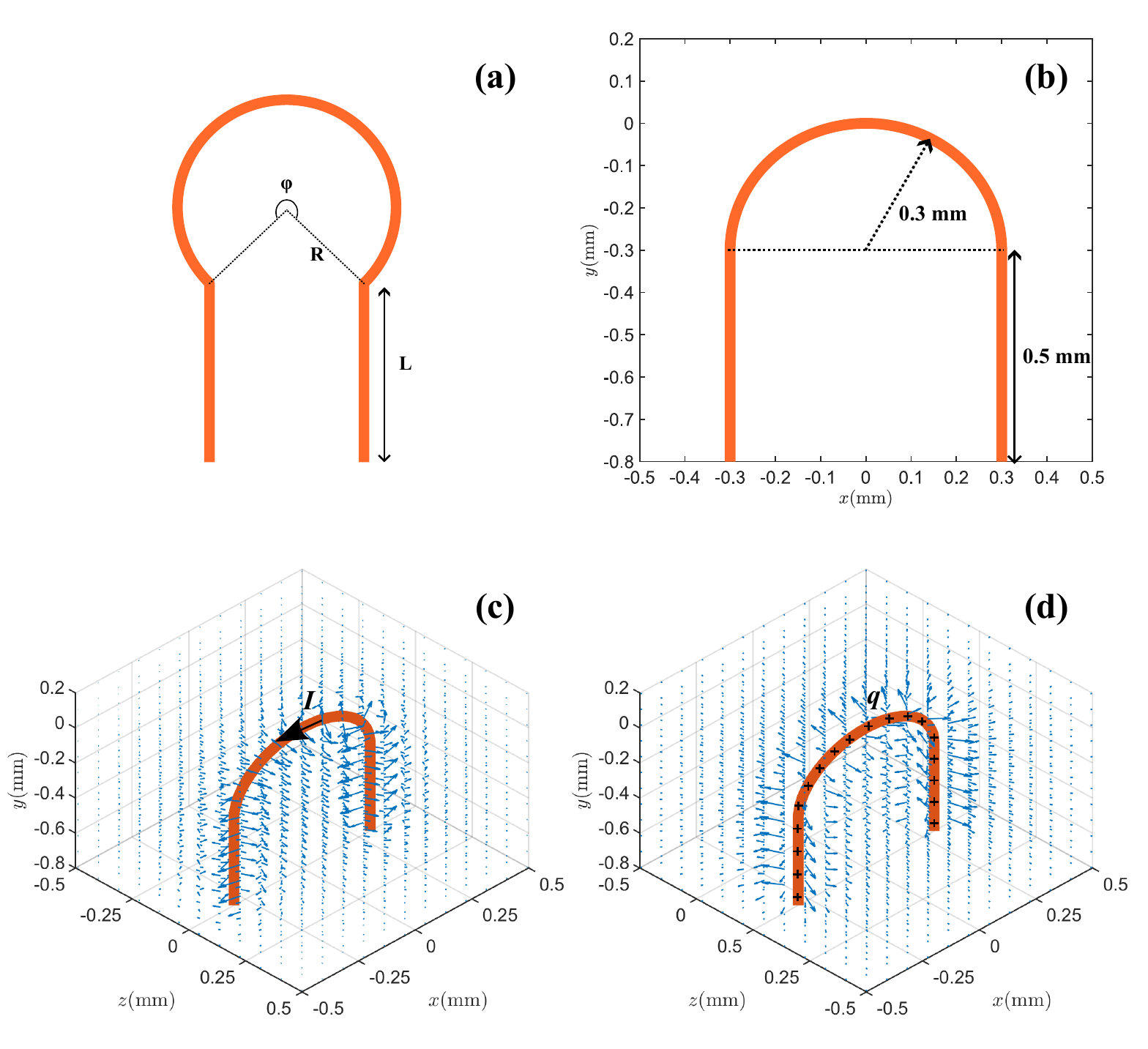}
\caption{(a) A general coil structure. It consists of two legs with a length L connected by a circular arc with a radius R and a central angle $\varphi$. (b) The dimensions of the coil used in this paper. The leg length $L=0.5$ mm, the circular arc has a radius $R=0.3$ mm and $\varphi=\pi$. 
(c) Magnetic field distribution generated by a current flowing through the coil, calculated using Equation \ref{B}. (d)  Electric field distribution produced by a uniform positive charge along the coil, calculated using Equation \ref{E}.}.
\label{fig:1}
\end{figure*}

We generate synthetic proton radiographs through particle tracing using the charged particle radiography module from PlasmaPy \cite{plasmapy_community_2024_12788848}. The primary inputs for this process are the electromagnetic fields and the relative positions of the proton source, target, and detector.

A general coil target is illustrated in Figure 1(a). The coil comprises two straight legs joined by a circular arc and is characterized by three parameters:  the leg length $L$, the radius of the circular arc radius $R$, and the central angle of the arc $\varphi$. In this paper, the chosen dimensions are $L=0.5$ mm, $R=0.3$ mm and $\varphi=\pi$, forming a U-shaped coil, as depicted in Figure \ref{fig:1}(b). This geometry is the same or similar as we used in previous experiments\cite{gao2016ultrafast}. The magnetic field is produced by the current flowing in a U-shaped coil. Magnetic field distribution is computed using the Biot-Savart law:
\begin{equation}
{\bf{B}} = \frac{{{\mu _0}I}}{{4\pi }}\int {\frac{{d{\bf{l'}} \times \left( {{\bf{r}} - {\bf{r'}}} \right)}}{{{{\left| {{\bf{r}} - {\bf{r'}}} \right|}^3}}}} .
\label{B}
\end{equation}
where $I$ is the current. 
Figure \ref{fig:1}(c) presents the three-dimensional magnetic field distribution near the coil, showing magnetic field lines circulating around the coil.
The electric field is generated by a U-shaped charge line along the coil, with a line charge density denoted as $\lambda$.  The electric field is calculated from Coulomb's law
\begin{equation}
{\bf{E}} = \frac{1}{{4\pi {\varepsilon _0}}}\int {\frac{{\lambda ds}}{{{{\left| {{\bf{r}} - {\bf{r'}}} \right|}^3}}}\left( {{\bf{r}} - {\bf{r'}}} \right)} .
\label{E}
\end{equation}
Figure \ref{fig:1}(d) displays the three-dimensional electric field distribution resulting from a positive charge distribution, with electric field lines radiating outward from the coil. For a negative charge distribution, the electric field lines reverse direction, converging inward toward the coil.

\begin{figure*}[h]
\centering

\includegraphics[width=0.8\textwidth]{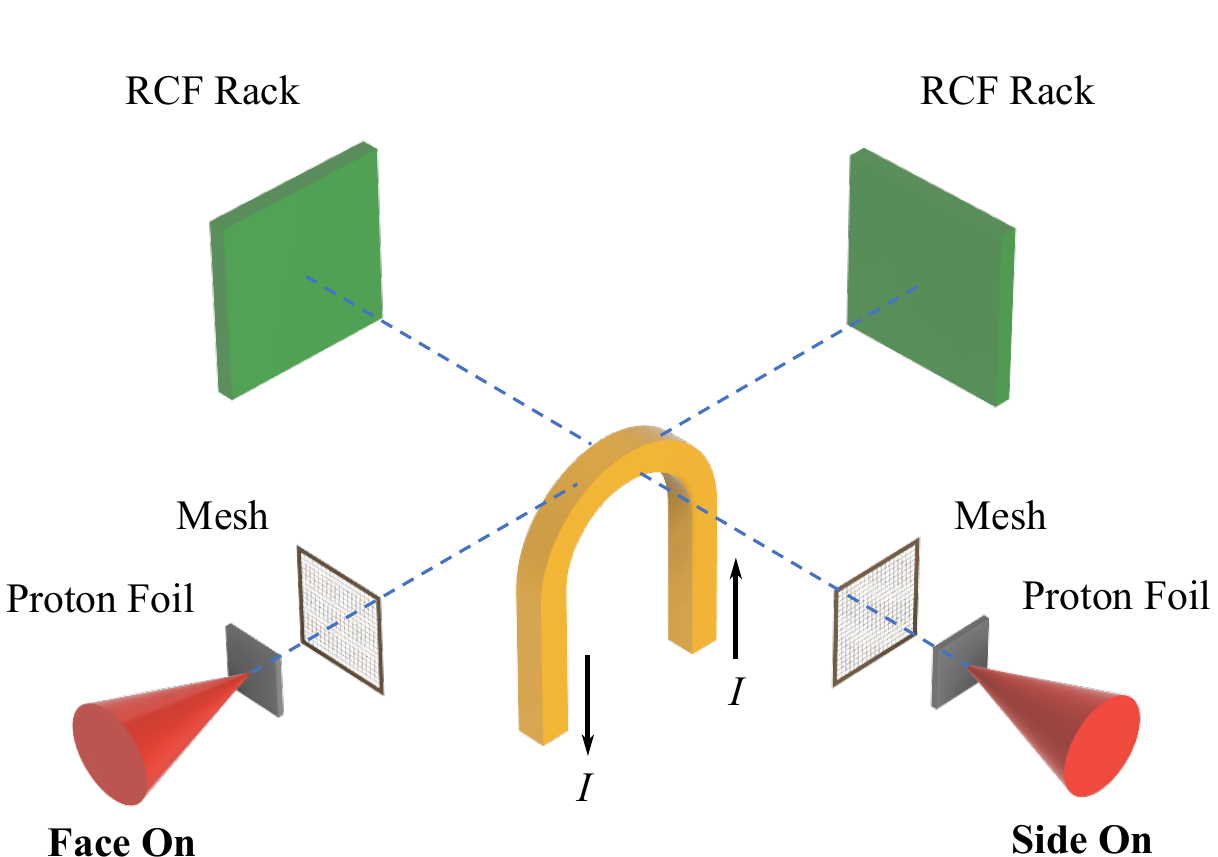}
\caption{Schematic representation of the proton radiography setup, in both face-on and side-on directions. The proton source is 7 mm away from the coil and the RCF rack is 80 mm away from the coil. A mesh is placed between the coil and foil at a distance 2.5 mm to the proton source. The mesh has a spacing of 53.57 $\mu$m, corresponding to a 150 $\mu$m spacing at the coil plane.}%
\label{fig:2}
\end{figure*}

The schematic for the proton radiography setup is shown in Figure \ref{fig:2}. The proton source is aligned with the center of the top of the coil along two different axes: face-on (+x-axis) and side-on (-z-axis). The source is positioned 7 mm from the target, and the image plane, represented here as a RCF rack is located 80 mm from the target. A mesh is placed between the proton source and the target, 2.5 mm from the source. The particles are removed when they hit the mesh. The protons are emitted from a point source, similar to those generated by the target-normal sheath acceleration mechanism when a high power laser irradiate a foil\cite{Gao2012}, as is shown in Figure \ref{fig:2}. The proton energy used in the simulation is 24.7 MeV. 

\section{Results}
Synthetic proton radiographs are generated for both face-on and side-on configurations, and the resulting structural features are systematically analyzed. The sharp boundaries observed in the features are attributed to caustics, which result from abrupt changes in proton accumulation\cite{gao2016ultrafast}. Note that proton radiographs are generally analyzed through the standard method by solving an inverse problem \cite{kugland2012invited}, however that method cannot be used for the significant deflections considered here. This necessitates the use of forward modeling used in this work. To facilitate a clear understanding of the origins of the caustic structures, the proton radiographs are calibrated from the image plane to the target plane.

\subsection{Face on proton radiographs}

\begin{figure*}[h]
\centering
\includegraphics[width=1\textwidth]{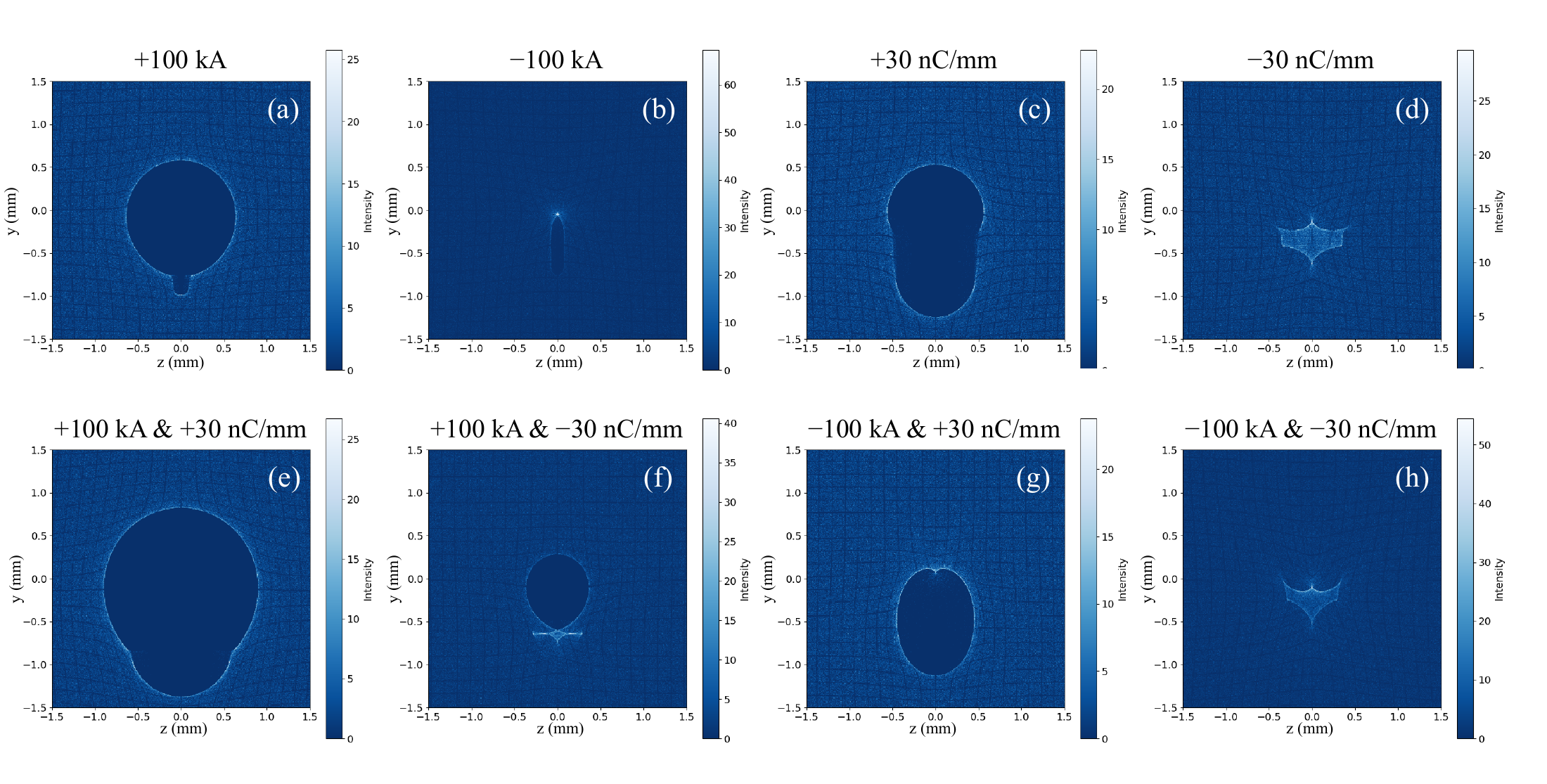}
\caption{ 
Face-on proton radiographs under various magnetic and electric field configurations. The magnetic fields arise from a $\pm$100 kA coil current, while the electric fields originate from a line charge with a density of $\pm$30 nC/mm. Each radiograph illustrates the distinct proton deflection patterns resulting from different combinations of these fields. The specific coil current and charge density values are indicated for each image.
 }
\label{fig:3}
\end{figure*}

Figure \ref{fig:3}(a) shows the proton radiograph generated from protons passing through the magnetic field produced by a $+$100 kA current in face-on configuration.  The radiograph reveals a central proton void bubble structure and a distinct tip feature at the base of the leg. The grid pattern expands outward from the center.
The trajectory of protons can be explained by the electromotive force $q{\bf{v}} \times {\bf{B}}$ and can also be understood as the interaction between two currents: two same-direction currents attract each other and two reverse-direction currents repel each other.
The proton beam can be thought of as a +$x$-directed current while the current at the top of the coil is oppositely directed, therefore protons are deflected away from the coil and form a bubble-shaped void. At the foot of the leg, the proton beam has an $-y$-directed current, so the first leg ($+y$-directed current) focuses the proton beam, and the second leg ($-y$-directed current) defocuses it, forming the extended deflection tip. Figure \ref{fig:3}(b) shows the proton radiograph for a reverse $-$100 kA current in the same configuration. It features a focused central point and an extended bar structure below. At the coil peak, the coil is of the same direction current as the proton beam and so the proton beam is attracted to the coil and forms a focus point. At the leg, the proton beam is defocused by the first leg and then focused by the second leg forming the extended deflection bar.
In contrast, Figure \ref{fig:3}(a) does not show the full deflection bar because the semicircular part of the coil causes stronger deflection than the upper leg, obscuring the bar structure at the leg region.

Figure \ref{fig:3}(c) presents the proton radiograph from protons passing through the electric field from a line charge density $+30\;{{{\rm{nC}}} \mathord{\left/
 {\vphantom {{{\rm{nC}}} {{\rm{mm}}}}} \right.
 \kern-\nulldelimiterspace} {{\rm{mm}}}}$. 
It shows a deflected structure with a top half structure similar to the magnetic field case and an elongated deflected void near the leg region showing the protons are deflected similarly at the top circle and the leg. The grid expands outward from the center similar to the magnetic field case. The leg deflection is comparable to the top deflection because electric field forces from charges are always along the electric field direction that is pointing away from the coil and are not related to the velocity of protons. For a negative line charge density of $-30\;{{{\rm{nC}}} \mathord{\left/
 {\vphantom {{{\rm{nC}}} {{\rm{mm}}}}} \right.
 \kern-\nulldelimiterspace} {{\rm{mm}}}}$, the proton radiograph in Figure \ref{fig:3}(d) shows a focusing symmetrical wing-like structures with protons attracted toward the coil centering at the connection point between the leg and the top semicircle. The grid contracts toward the coil, indicating that protons experience an attractive force as they traverse the negatively charged region.

When combining the electromagnetic fields of a $+$100 kA current and a  $+30\;{{{\rm{nC}}} \mathord{\left/
 {\vphantom {{{\rm{nC}}} {{\rm{mm}}}}} \right.
 \kern-\nulldelimiterspace} {{\rm{mm}}}}$ positive charge distribution, the radiograph reveals a larger void bubble and a more pronounced deflection tip as shown in Figure \ref{fig:3}(e).
With the combined fields of a $+100$ kA current and a $-30\;{{{\rm{nC}}} \mathord{\left/
 {\vphantom {{{\rm{nC}}} {{\rm{mm}}}}} \right.
 \kern-\nulldelimiterspace} {{\rm{mm}}}}$ negative charge distribution, the radiograph in Figure \ref{fig:3}(f) exhibits a smaller void bubble alongside an attractive feature at the leg's base. The grid expands at the top semicircle and contracts toward the leg, demonstrating the interplay of magnetic and electric forces in shaping proton trajectories. 
 
Under the electromagnetic fields generated by a reverse $-$100 kA current and a  $+30\;{{{\rm{nC}}} \mathord{\left/
 {\vphantom {{{\rm{nC}}} {{\rm{mm}}}}} \right.
 \kern-\nulldelimiterspace} {{\rm{mm}}}}$ positive charge distribution,  protons are focused at the peak of the coil, creating a small void bubble just below the focal point as shown in Figure \ref{fig:3}(g). The top half of the grid contracts inward toward the coil peak, while the bottom half expands outward from the bubble region.
With the combined fields of a reverse $-$100 kA current and a $-30\;{{{\rm{nC}}} \mathord{\left/
 {\vphantom {{{\rm{nC}}} {{\rm{mm}}}}} \right.
 \kern-\nulldelimiterspace} {{\rm{mm}}}}$ negative charge distribution, the radiograph in Figure \ref{fig:3}(h) exhibits a similar focusing symmetrical wing-like structures pattern to that in (d), with protons drawn toward the coil. The grid contracts inward toward the focus structure.

\subsection{Side on proton radiographs}

\begin{figure*}[h]
\centering
\includegraphics[width=1\textwidth]{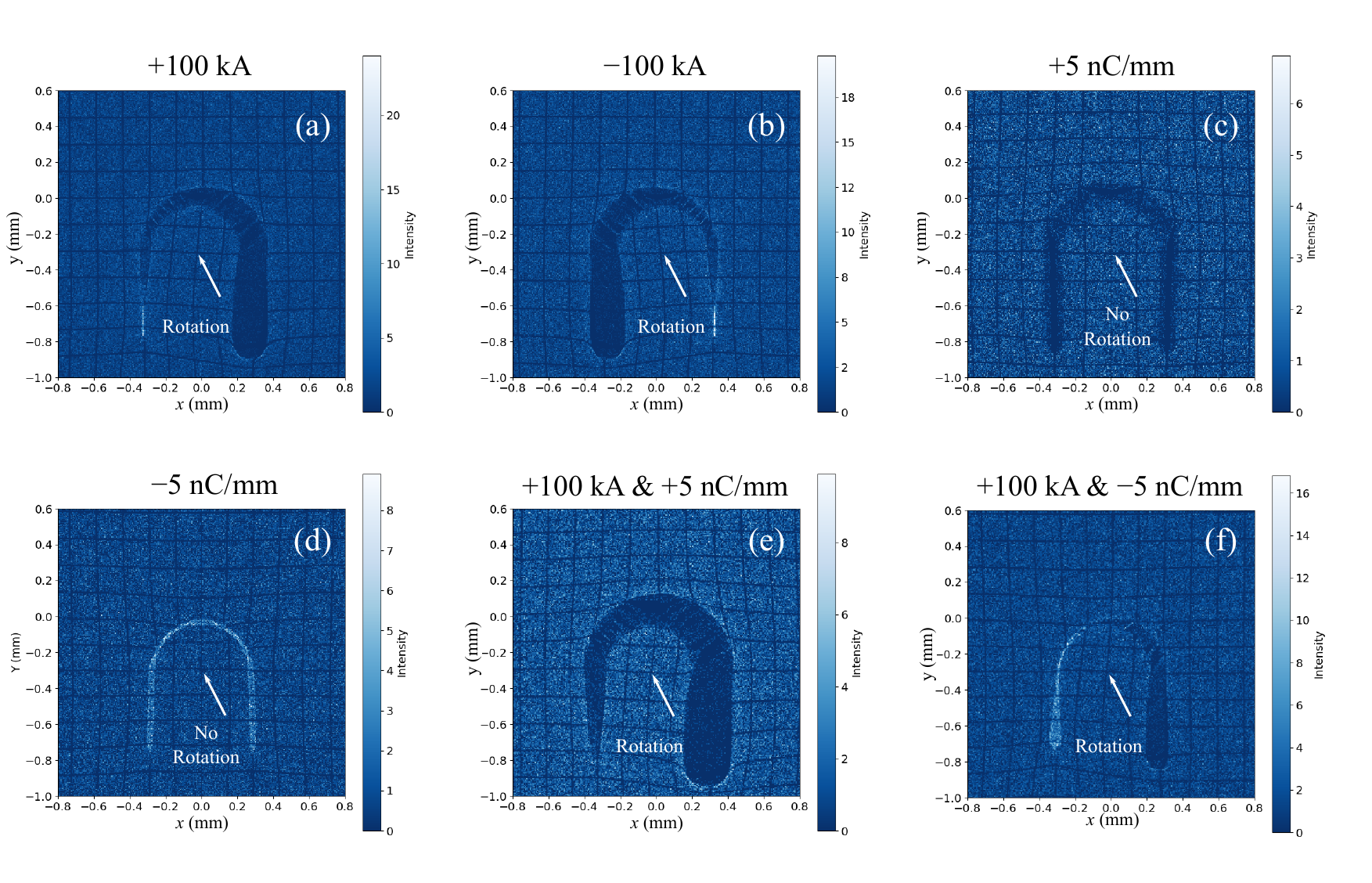}
\caption{ 
Side-on proton radiographs under various magnetic and electric field configurations. The magnetic fields arise from a $\pm$100 kA coil current, while the electric fields originate from a line charge with a density of $\pm$5 nC/mm.
Each radiograph reveals unique proton deflection patterns corresponding to various field combinations, with the specific coil current and charge density values labeled for each image.
}
\label{fig:4}
\end{figure*}

Figure \ref{fig:4}(a) shows the side-on proton radiograph for the magnetic field-only case, with a +100 kA current flowing through the coil. The proton patterns near the two legs exhibit asymmetry: the left leg appears narrow, while the right leg is wider. This difference arises from the opposite current directions in the two legs—downward in the left leg and upward in the right leg. The protons, which have a downward velocity component, are attracted by the left leg, creating a narrow feature, and repelled by the right leg, resulting in a wider feature. Another notable aspect is the rotation of the central grid. This rotation is caused by the shifting difference of protons at different heights. Proton beams passing through lower regions of the coil have a higher downward velocity or downward current, which leads to stronger deflection toward the left leg, shifting them further leftward.
Figure \ref{fig:4}(b) presents the side-on proton radiograph for the magnetic field-only case, where a reverse current of $-100$ kA is applied through the coil. The radiograph mirrors the pattern observed in Figure \ref{fig:4}(a).

Figure \ref{fig:4}(c) presents the side-on proton radiograph for the electric field-only case with a line charge density of $+5\;{{{\rm{nC}}} \mathord{\left/
 {\vphantom {{{\rm{nC}}} {{\rm{mm}}}}} \right.
 \kern-\nulldelimiterspace} {{\rm{mm}}}}$.  This lower charge density is chosen to produce a proton deflection comparable to that in the magnetic field-only case shown in Figure \ref{fig:4}(a). It has a deflection pattern with nearly uniform widths along the coil, as the electric field repulsion is nearly symmetric along the coil. Consequently, the two legs appear symmetric. Additionally, there is no central grid rotation because protons passing through the center experience equal repelling forces from both sides. Near the coil, the grids are pushed outward due to electrostatic repulsion. Figure \ref{fig:4}(d) presents the side-on proton radiograph for the electric field-only case with a line charge density of $-5\;{{{\rm{nC}}} \mathord{\left/
 {\vphantom {{{\rm{nC}}} {{\rm{mm}}}}} \right.
 \kern-\nulldelimiterspace} {{\rm{mm}}}}$. It exhibits a symmetric attraction pattern along the coil. The electric field attraction creates nearly uniform widths for both legs, making them appear symmetric. Similar to the positively charged case, no central grid rotation occurs because protons passing through the center experience equal attracting forces from both sides. Near the coil, the grids are pulled inward due to electrostatic attraction.
 
Figure \ref{fig:4}(e) shows the side-on proton radiograph for the combined magnetic and electric field case, with a +100 kA current and a line charge density of $+5\;{{{\rm{nC}}} \mathord{\left/
 {\vphantom {{{\rm{nC}}} {{\rm{mm}}}}} \right.
 \kern-\nulldelimiterspace} {{\rm{mm}}}}$. It retains the asymmetric leg feature observed in the magnetic field-only case but with wider leg patterns. The central grid rotation remains the same as that from the magnetic field only case, indicating that the rotation angle can be used as a diagnostic tool for determining the magnetic field generation. Figure \ref{fig:4}(f) shows the side-on proton radiograph for the combined magnetic and electric field case, with a +100 kA current and a line charge density of $-5\;{{{\rm{nC}}} \mathord{\left/
 {\vphantom {{{\rm{nC}}} {{\rm{mm}}}}} \right.
 \kern-\nulldelimiterspace} {{\rm{mm}}}}$. The radiograph again retains the asymmetric leg feature but with narrower leg patterns compared to the magnetic field-only case. The central grid rotation persists, confirming its reliability in identifying the magnetic field generation from the coil. The combined fields cases with -100 kA coil current are mirroring the features in (e) and (f) and are not shown here. 

\renewcommand{\figurename}{Figure}
\setcounter{figure}{4}
\begin{figure*}[h]
\centering
\includegraphics[width=0.6\textwidth]{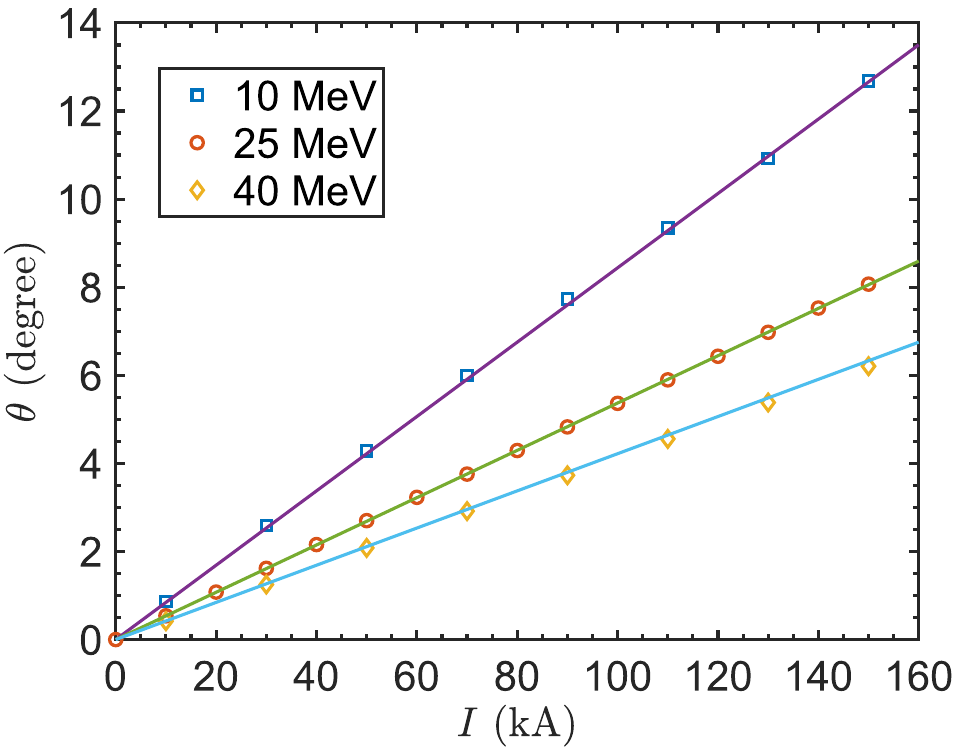}
\caption{The central grid rotation angle from side-on proton radiographs versus coil current at proton energies $E_p$ = 10, 25 and 40 MeV. The scattered points represent simulation results, while the lines are fitted using Equation \ref{theta} with $\alpha_{eff}$ = 0.27.
}
\label{fig:5}
\end{figure*}

The central grid rotation angle from side-on proton radiography can be analytically estimated. Considering that particles pass through the central grid region with an initial velocity in the $y$ and $z$ directions. $v_y \ll v_z$ and energy $E_p \approx \frac{1}{2}mv_z^2$, the $x$ directional velocity at the detector is 
\begin{equation}
{v_x} \approx \frac{{q{v_y}}}{{\sqrt {m{E_p}} }}\int {\left( {{B_z} - \frac{{{v_z}}}{{{v_y}}}{B_y}} \right)dz}.   
\label{v_x}
\end{equation}
The integral has a dimension proportional to ${\mu _0}I$ and can be expressed as $\int {\left( {{B_z} - \frac{{{v_z}}}{{{v_y}}}{B_y}} \right)dz} =\alpha{\mu _0}I$, where $\alpha$ is a coefficient determined from simulations. Substituting this into the expression for $v_x$, we have
\begin{equation}
{v_x} = \alpha \frac{{{\mu _0}qI{v_y}}}{{\sqrt {m{E_p}} }}.   
\end{equation}
The rotation angle can be calculated as
\begin{equation}
\theta=\arctan \left( {\frac{x}{y}} \right) \approx \arctan \left( {\frac{v_x}{v_y}} \right)   
\end{equation}
For a small rotation angle, $\arctan \left( {\frac{x}{y}} \right) \approx \frac{{{v_x}}}{{{v_y}}}$, leading to:
\begin{equation}
\theta  \approx \alpha \frac{{{\mu _0}q}}{\sqrt{m}}\frac{I}{{\sqrt {{E_p}} }}.  
\end{equation}
This equation can be expressed in practical units as 
\begin{equation}
\theta \left[ ^\circ  \right] = {\alpha _{eff}}\frac{{I\left[ {{\rm{kA}}} \right]}}{{\sqrt {{E_p}\left[ {{\rm{MeV}}} \right]} }}.  
\label{theta}
\end{equation}
where $\alpha_{eff}$ is a dimensionless coefficient and can be determined from simulations. 
Figure 5 shows the central grid rotation angle as a function of coil current for  $E_p=$ 10, 25 and 40 MeV. The rotation angle is defined over the central region of the coil—from $y = 0.15~\mathrm{mm}$ to $y = 0.45~\mathrm{mm}$, centered at $y = 0.30~\mathrm{mm}$—where the rotation angle remains relatively constant.
The scattered points represent simulation data, while the lines correspond to fits based on Equation \ref{theta}. The coefficient 
$\alpha_{eff}=0.27$ is obtained by fitting the 25 MeV data and is used consistently for all three curves. The results demonstrate that Equation \ref{theta} accurately describes the relationship between rotation angle and current.
In practical applications, the current is estimated from the measured rotation angle. Rearranging the equation, the current for the coil can be estimated as:
\begin{equation}
I\left[ {{\rm{kA}}} \right] \approx 3.7\sqrt {{E_p}\left[ {{\rm{MeV}}} \right]} \theta \left[ ^\circ  \right].  
\label{I}
\end{equation}
For coil targets with different geometries, the numerical coefficients in Eqs. \ref{theta} and \ref{I} may change. These coefficients can be recalculated using the same approach described here, tailored to the specific geometry and magnetic field profile.

\subsection{Determining the electromagnetic fields}
Figure \ref{table:proton_radiographs} presents a summary table of the proton radiograph features for different cases in face-on and side-on proton radiography configurations. From the features, one can qualitatively determine the dominant field contribution in an experimental proton radiograph.    

\subsubsection{Side-On Grid Rotation Method}
\hfill \break

For a general quantitative measurement of the electromagnetic field, side-on proton radiography with a mesh is recommended. 

The grid rotation, attributed exclusively to the magnetic field, serves as a reliable indicator for determining current and magnetic field strength. Moreover, the rotation remains independent of magnification, which is defined as the ratio between the distance from the proton source to the detector and the distance from the proton source to the target. This independence helps minimize position-related errors from the position of the proton source, the target, and the RCF rack when measuring feature sizes. For instance, in a face-on measurement, the current is determined by the size of the bubble, which must be recalibrated to the target plane using the magnification factor. However, in real experiments, inaccuracies in positioning the proton source and the RCF rack can introduce errors in size calculations, subsequently affecting the current measurement. In contrast, the angle is calculated as a ratio of two lengths, making it inherently independent of the magnification factor and therefore more robust to positional errors.

Here are the two steps to apply the method: (1) Determine the magnetic field and current from the central grid rotation angle from Equation \ref{I}. (2) Once the magnetic field is estimated from the grid rotation, proton radiography calculations can be conducted to compare the leg width with experimental results. Any observed differences in leg width can then be attributed to the contribution of the electric field. 

To apply the grid rotation method, a sufficiently large rotation angle is required—typically at least 1 degree for reliable measurement based on our experiment. This threshold may vary across different experimental setups, as the minimum resolvable rotation angle depends on factors such as proton energy, coil geometry, and the relative positions of the proton source, coil, and detector. For the coil geometry discussed in this paper, the minimum measurable current under 1 degree rotation must satisfy
$I~[\mathrm{kA}] > 3.7\sqrt{E_p~[\mathrm{MeV}]},$
where $E_p$ is the proton beam energy. For lower current scenarios, using a lower-energy proton beam is feasible. 

\subsubsection{Face-On Bubble and Leg Method}
\hfill \break

For face-on proton radiography, a positive current configuration—where the proton beam moves in the opposite direction to the coil current—is preferable. In this setup, the bubble void width at the circular arc is directly related to the current and charge density amplitude, as described in \cite{gao2016ultrafast, bradford2021measuring}.

The distinct patterns at the circular arc and the leg provide valuable insights into the relative contributions of electric and magnetic fields to the bubble size. By simulating proton radiography and fitting both the central bubble and the extended structure at the leg, the current and distributed charge density can be determined.

\subsubsection{Application to experiments}
\hfill \break

Our approach introduces two complementary diagnostic methods that extract both magnetic and electric field components from a single radiograph, in either side-on or face-on geometry. Each method leverages two independent and physically distinct observables, forming a coupled system of equations that can be uniquely solved. This strategy eliminates the need for multiple shots or broad proton energy spectra, while providing quantitative access to both magnetic and electric field contributions in a single experiment.

Using the side-on grid rotation method, we successfully measured a 120 kA current flowing through a double-coil target driven by a short-pulse laser \cite{ji2024study,gao2025record}. 
This current generates a 200 T magnetic field at the center of one circular arc. When we overlay the simulation results for the 120 kA current onto the experimental radiograph, we observe excellent agreement in both grid rotation and leg widths.
To estimate the upper-bound charge density and electric field, we included an additional electric field in the simulation. The resulting upper-bound charge density is $\pm 0.5\;{{{\rm{nC}}} \mathord{\left/{\vphantom {{{\rm{nC}}} {{\rm{mm}}}}} \right.\kern-\nulldelimiterspace} {{\rm{mm}}}}$, corresponding to an upper-bound electric field strength of approximately $3 \times 10^{7}$ V/m at a distance of 0.3 mm away from the coil.  Additionally, this platform for generating strong magnetic fields using short-pulse lasers is also applied to magnetic reconnection studies. Strong particle acceleration is observed under this intense magnetic field. Further details can be found in another paper\cite{gao2025record}.

\subsubsection{Comparison of Electric and Magnetic Deflection Contributions}
\hfill \break

The characteristic field strengths scale as:
$B \sim \frac{\mu_0 I}{2\pi r}$, $\quad E \sim \frac{\lambda}{2\pi \epsilon_0 r}$, where $I$ is the coil current, $\lambda$ is the line charge density, and $r$ is the distance from the source. The ratio of electric to magnetic deflection for a beam with velocity $v_\perp$ orthogonal to the magnetic field direction can then be estimated as: 
\begin{equation}
\frac{qE}{qv_\perp B} \sim \frac{\lambda c^2}{I v_{\perp}}  
\label{I2}.
\end{equation}
Here, $v_\perp=v_z$ in face-on configuration and $v_\perp=v_y$ in side-on direction.

This expression indicates that when $\lambda/I \ll v_\perp/c^2$, the magnetic deflection dominates; whereas if $\lambda/I \gg v_\perp/c^2$, the electric deflection becomes dominant. This estimate provides a practical method for determining the electric and magnetic contributions from line charge density and current.

\section{Conclusions}
\begin{figure}[ht]
\includegraphics[width=1\textwidth]{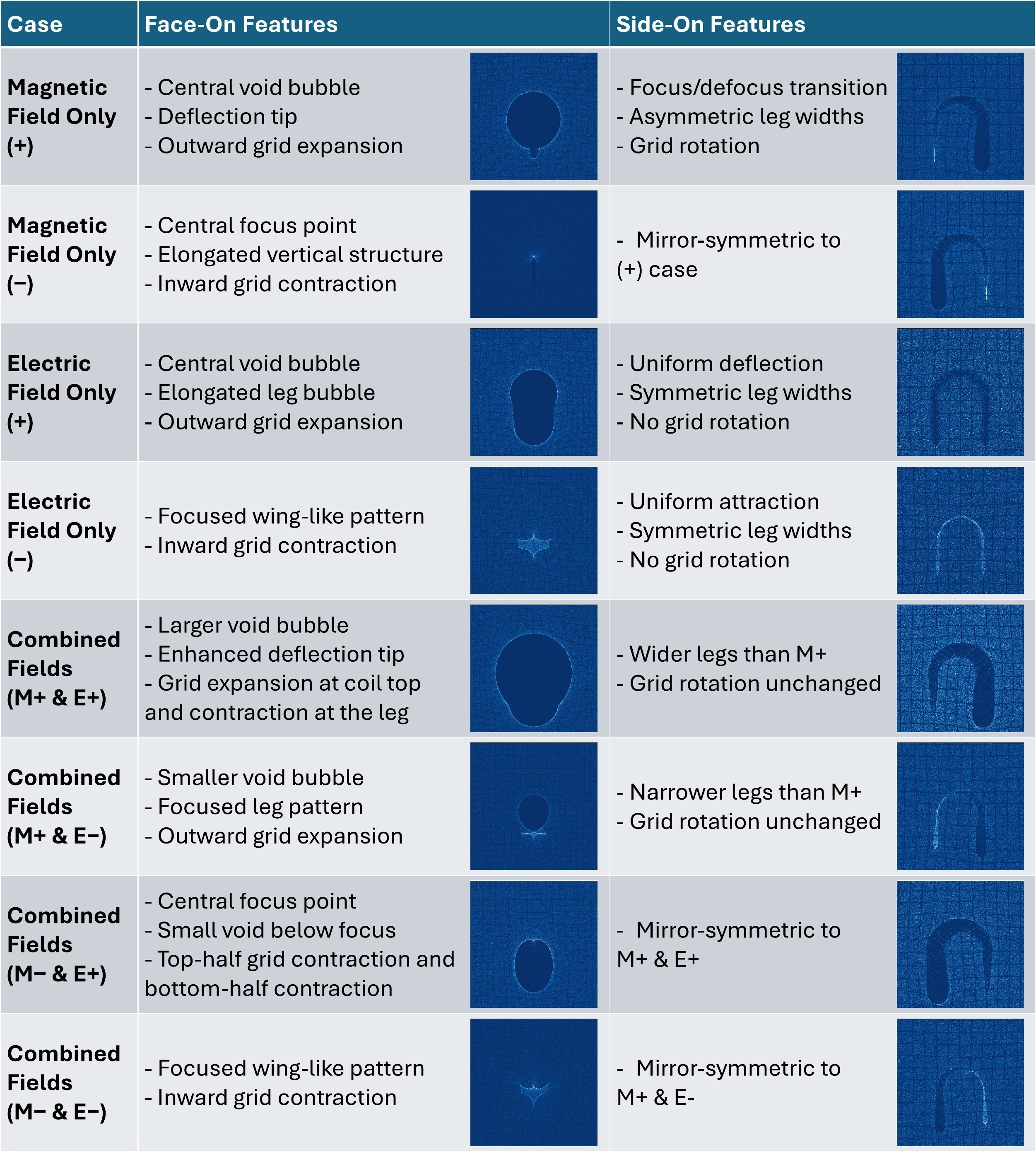}
\caption{Comparison table of proton radiograph features for various cases in face-on and side-on configurations. "$+$" denotes positive current or charges as defined in Figure \ref{fig:2}, while "$-$" represents negative current or charges.}%
\label{table:proton_radiographs}
\end{figure}

In summary, we systematically investigated the features of proton radiographs under varying electromagnetic field conditions using face-on and side-on configurations. By analyzing the distinct proton deflection patterns in these configurations, we identified key features attributable to magnetic and electric fields, as summarized in Figure \ref{table:proton_radiographs}. Based on these findings, we propose methods to isolate and quantify the magnetic field contribution using the side-on central grid rotation and the face-on bubble and leg width analysis. The side-on central grid rotation method is particularly preferable, as the rotation is uniquely sensitive to the magnetic field.  The electric field contribution can then be determined by comparing leg width differences between combined-field and magnetic-field-only cases. We applied the side-on central grid rotation method to evaluate the electromagnetic fields generated by a U-shaped double-coil target, demonstrating its effectiveness in enhancing the diagnostic capabilities of proton radiography for accurate and reliable field measurements in high-energy-density experiments.  Moreover, the detailed feature analysis presented in this work provides a better understanding of the interplay between electric and magnetic field contributions in proton radiographs, facilitating the optimization of laser-driven capacitor coil designs for advanced applications. Looking ahead, this proton radiograph analysis could be extended to more complex field geometries, including those with non-uniform, dynamic, or time-varying current and charge distributions. 

\section{Acknowledgements}
This work was supported by US Department of Energy the High-Energy-Density Laboratory Plasma Science program under Grant No. DE-SC0020103, the NASA Living with a Star Jack Eddy Postdoctoral Fellowship Program administered by UCAR's Cooperative Programs for the Advancement of Earth System Science (CPAESS) under award $\#$80NSSC22M0097, the LaserNetUS initiative at the OMEGA EP Laser System, and the Laboratory Basic Science program at the Laboratory for Laser Energetics. Yang Zhang thanks Peter Heuer for helpful discussions in using the radiography module from PlasmaPy.

\section{References}
\bibliographystyle{unsrt}

\end{document}